\documentclass[sigconf]{acmart}

\usepackage{subfig}%多个子图
\usepackage{amsmath}
\usepackage{enumitem} 
\usepackage{multirow}
\usepackage{threeparttable}
\let\oldhat\hat
\renewcommand{\vec}[1]{\mathbf{#1}}
\renewcommand{\hat}[1]{\oldhat{\mathbf{#1}}}
\renewcommand{\matrix}[1]{\mathbf{#1}}

\newcommand{\ie}{\emph{i.e., }}
\newcommand{\eg}{\emph{e.g., }}

\newcommand{\paratitle}[1]{\vspace{1ex}\noindent \textbf{#1}}

\AtBeginDocument{%
  }

\setcopyright{acmlicensed}
\copyrightyear{2018}
\acmYear{2018}
\acmDOI{XXXXXXX.XXXXXXX}

\acmConference[Conference acronym 'XX]{Make sure to enter the correct
  conference title from your rights confirmation emai}{June 03--05,
  2018}{Woodstock, NY}
\acmISBN{978-1-4503-XXXX-X/18/06}

\begin{document}

%%
%% The "title" command has an optional parameter,
%% allowing the author to define a "short title" to be used in page headers.
\title{Simple but Efficient: A Multi-Scenario Nearline Retrieval Framework for Recommendation on Taobao}

\author{Yingcai Ma}
\email{yingcai.myc@taobao.com}
\affiliation{%
  \institution{Alibaba Group}
  \country{China}
}

\author{Ziyang Wang}
\email{shanyi.wzy@taobao.com}
\affiliation{%
  \institution{Alibaba Group}
  \country{China}
}

\author{Yuliang Yan}
\email{yuliang.yyl@alibaba-inc.com}
\affiliation{%
  \institution{Alibaba Group}
  \country{China}
}

\author{Jian Wu}
\email{joshuawu.wujian@taobao.com}
\affiliation{%
  \institution{Alibaba Group}
  \country{China}
}

\author{Yuning Jiang}
\email{mengzhu.jyn@taobao.com}
\affiliation{%
  \institution{Alibaba Group}
  \country{China}
}

\author{Longbin Li}
\email{lilongbin.llb@taobao.com}
\affiliation{%
  \institution{Alibaba Group}
  \country{China}
}

\author{Wen Chen}
\email{chenyu.cw@taobao.com}
\affiliation{%
  \institution{Alibaba Group}
  \country{China}
}

\author{Jianhang Huang}
\email{jianhang.hjh@alibaba-inc.com}
\affiliation{%
  \institution{Alibaba Group}
  \country{China}
}

%%
%% The "author" command and its associated commands are used to define
%% the authors and their affiliations.
%% Of note is the shared affiliation of the first two authors, and the
%% "authornote" and "authornotemark" commands
%% used to denote shared contribution to the research.

%%
%% By default, the full list of authors will be used in the page
%% headers. Often, this list is too long, and will overlap
%% other information printed in the page headers. This command allows
%% the author to define a more concise list
%% of authors' names for this purpose.
% \renewcommand{\shortauthors}{Trovato et al.}

%%
%% The abstract is a short summary of the work to be presented in the
%% article.
\begin{abstract}
    In recommendation systems, the matching stage is becoming increasingly critical, serving as the upper limit for the entire recommendation process. Recently, some studies have started to explore the use of multi-scenario information for recommendations, such as model-based and data-based approaches. However, the matching stage faces significant challenges due to the need for ultra-large-scale retrieval and meeting low latency requirements. As a result, the methods applied at this stage~(\eg collaborative filtering and two-tower models) are often designed to be lightweight, hindering the full utilization of extensive information. On the other hand, the ranking stage features the most sophisticated models with the strongest scoring capabilities, but due to the limited screen size of mobile devices, most of the ranked results may not gain exposure or be displayed.
    
    In this paper, we introduce an innovative multi-scenario nearline retrieval framework. It operates by harnessing ranking logs from various scenarios through Flink, allowing us to incorporate finely ranked results from other scenarios into our matching stage in near real-time. Besides, we propose a streaming scoring module, which selects a crucial subset from the candidate pool. Implemented on the "Guess You Like" (\ie homepage of the Taobao APP), China's premier e-commerce platform, our method has shown substantial improvements—most notably, a 5\% uptick in product transactions. Furthermore, the proposed approach is not only model-free but also highly efficient, suggesting it can be quickly implemented in diverse scenarios and demonstrate promising performance.
\end{abstract}

%%
%% The code below is generated by the tool at http://dl.acm.org/ccs.cfm.
%% Please copy and paste the code instead of the example below.
%%
\begin{CCSXML}
<ccs2012>
   <concept>
       <concept_id>10002951.10003317.10003347.10003350</concept_id>
       <concept_desc>Information systems~Recommender systems</concept_desc>
       <concept_significance>500</concept_significance>
       </concept>
 </ccs2012>
\end{CCSXML}

\ccsdesc[500]{Information systems~Recommender systems}
%%
%% Keywords. The author(s) should pick words that accurately describe
%% the work being presented. Separate the keywords with commas.
\keywords{Recommendation Systems, Large-scale Recommendation, Matching Algorithm}
%% A "teaser" image appears between the author and affiliation
%% information and the body of the document, and typically spans the
%% page.

%%
%% This command processes the author and affiliation and title
%% information and builds the first part of the formatted document.
\maketitle

\section{Introduction \& Related Work}

With the rapid advancement of e-commerce platforms, recommendation systems~\cite{kabbur2013fism,hsieh2017collaborative,ko2022survey} have become vital for personalized content recommendation, improving user experiences, and boosting business revenues.
In most industrial scenarios~\cite{gong2020edgerec,cai2023reinforcing}, recommendation systems are challenged with vast candidates and strict latency constraints, which hence consist of several stages~(\eg matching~\cite{xie2021improving}, pre-ranking~\cite{zhang2023rethinking} and ranking~\cite{xia2008listwise}) to balance efficiency and accuracy.
The matching stage is designed to quickly narrow down relevant items from a vast pool of candidates, prioritizing low computational complexity for scalability. This sets the stage for the subsequent pre-ranking and ranking phases, where more granular scoring takes place building on the initial filtering.

\begin{figure}[t]
  \centering
  \small
  \includegraphics[width=\linewidth]{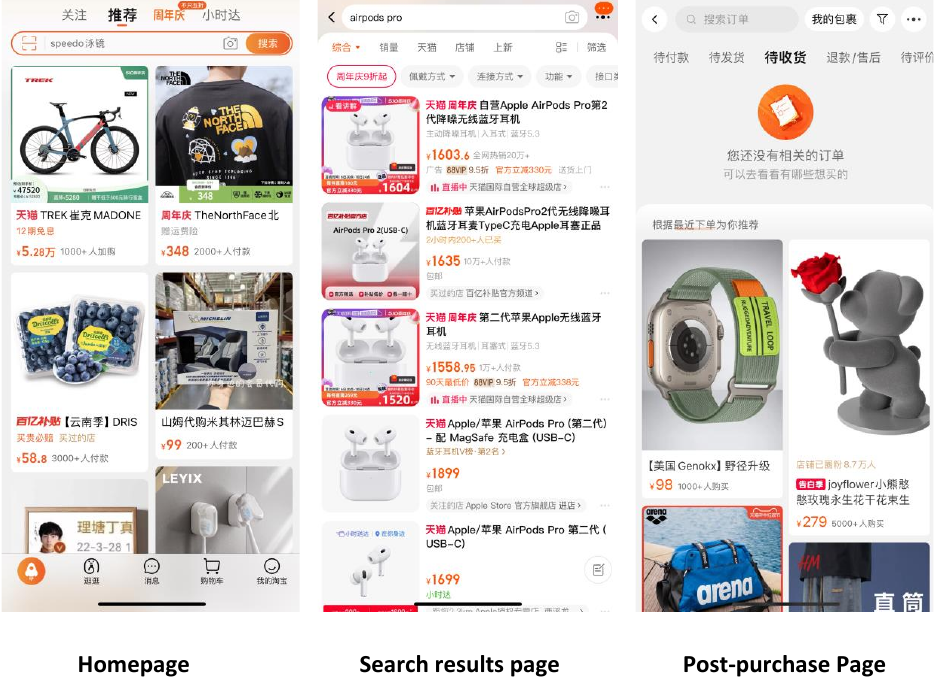}
  \vspace{-0.1cm}
  \caption{
    Different scenarios at Taobao.
  }
  \label{fig:scenePage}
  \vspace{-0.4cm}
\end{figure}

Positioned at the very outset of the recommendation systems, the matching phase inherently determines the maximum potential efficacy of recommendation quality.
Existing matching algorithms in recommendation systems predominantly fall under two broad categories: collaborative filtering-based methods~\cite{sarwar2001item,koren2009matrix,zheng2016neural} and deep learning-based methods~\cite{cheng2016wide,guo2017deepfm,lv2019sdm,cen2020controllable}. Collaborative filtering-based methods leverage historical interactions between users and items or latent factors to understand user preferences, yet they often grapple with issues related to data sparsity. On the other hand, deep learning-based approaches~\cite{zhang2019deep} typically make use of comprehensive neural network models in an end-to-end fashion to deduce how users might feel about each item. Although effective, these methods rely on simpler model structures~\cite{huang2013learning}, which might not fully grasp the complexities of user preferences. To address the limitations of model complexity in the matching phase, Netflix proposed a nearline retrieval framework~\cite{amatriain2013system} that utilizes indirect calculations for complex model inference. However, this approach suffers from delayed real-time responses and significantly increases resource expenditure.

Recently the concept of multi-scenario modeling~\cite{lu2016hierarchical,sheng2021one,tang2020progressive} has been introduced to recommendation systems. Using Taobao as a case in point, Figure \ref{fig:scenePage} illustrates various personalized recommendation scenarios including the homepage, search results page, and post-purchase page. User behavior observed in one scenario can potentially inform and enhance recommendations in another. Nevertheless, most current multi-scenario modeling approaches~\cite{zhu2019aligning,chang2023pepnet} employ complex frameworks, leading to a significant uptick in computational cost, rendering them impractical for the matching phase. In practice, within each scenario, the ranking stage evaluates thousands of candidates; however, due to page limitations, typically only tens of items are ultimately displayed to the user. As a result, a substantial number of high-scoring items from the ranking stage remain unseen by users despite their latent potential.
% Therefore, it becomes imperative to devise a framework that can identify and harness the potential of these high-scoring but underexposed products, ensuring that they receive the visibility they deserve and thereby improving the overall user experience and engagement.

% it is non-trivial to define an appropriate multi-scenario framework in matching stage.
% which requires us to address the following fundamental issues:
% (1) How to accurately learn the user’s preferences in other scenarios in the current matching stage? Directly training a recall model of the current context with data from a different context can be challenging due to the complexity constraints of the recall model, making it difficult to accurately capture user preferences from other contexts.
% (2) What methodologies can be employed to ascertain the continuous contemporaneity of user preferences? Given the inherently dynamic nature of user preferences, which are subject to frequent modifications, maintaining the real-time relevance of these preferences is of paramount importance in ensuring an accurate and up-to-date representation of user interests.

To address the above challenges, our paper presents a novel \emph{\textbf{M}ulti-scenario \textbf{N}earline \textbf{R}etrieval}~(MNR) approach for the matching stage. We leverage the power of complex ranking models from various scenarios by aggregating the ranking results for each user into a consolidated candidate pool. We then introduce a real-time scoring module that operates online, assessing the pool by considering both the initial ranking and the access time of each item to identify the optimal set for matching.
Implemented on Taobao, China's leading e-commerce platform, our approach has yielded significant performance gains in live environments, signaling a major advancement in the field of recommendation systems.

The major contributions are summarized as follows,
\begin{itemize}[leftmargin=*]
\item We exploit a multi-scenario nearline retrieval framework in the matching stage, which is both model-free and efficient. To the best of our knowledge, we are the first to harness the capabilities of complex ranking models across different scenarios for the matching stage.
\item The proposed MNR performs a streaming scoring module, which evaluates the candidate pool by considering both the initial order and access time of each item.
\item We carried out online experiments on Taobao, one of the globe's most expansive recommendation platforms, and observed a 5\% uptick in transaction volume. These findings underscore the efficiency and superiority of the framework we've introduced.
\end{itemize}

\section{Method}

\begin{figure*}[!htp]
  \centering
  \small
  \includegraphics[width=0.95\linewidth]{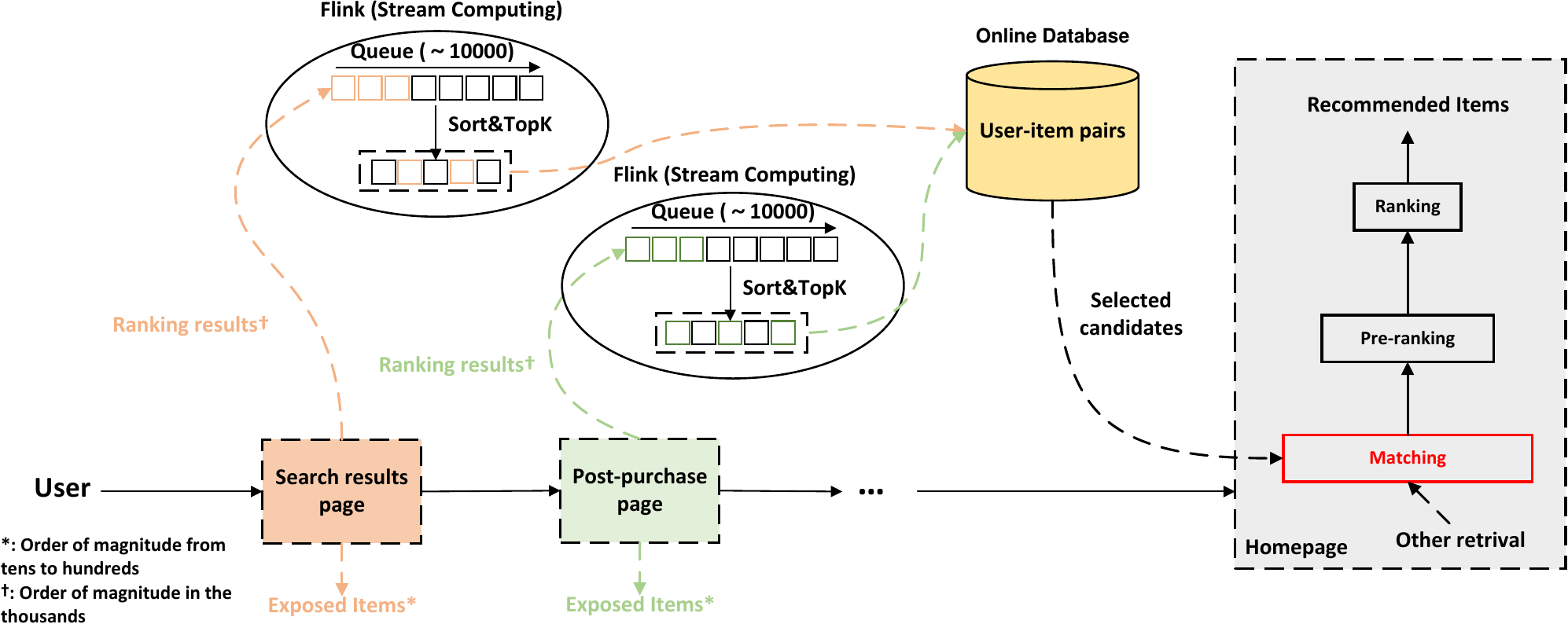}
  \caption{
    Multi-Scenarios Nearline Retrieval Framework on Taobao. As users engage with different scen arios, the fine-ranked outcomes are queued for processing. They are then scored and truncated via Flink, before being stored in an online database. Whenever users visit the homepage, the relevant data is fetched from the online database and fed into the matching process.
  }
  \label{fig:framework}  
  \vspace{-0.4cm}
\end{figure*}

\subsection{Problem Formalization}
The goal of the matching stage in industrial recommendation systems is to sift through a vast item pool $\matrix{E} = \{ \vec{e}_1, \vec{e}_2, \cdots, \vec{e}_{|E|} \}$, numbering in the billions, for each user $\vec{u}$. This process aims to narrow down the items to a tailored subset $\mathcal{E}_{u}$ ($|\mathcal{E}_{u}| << |\matrix{E}|$), ensuring that each item within this subset aligns with the user's interests,
\begin{equation}
      \mathcal{E}_{u} = \mathcal{F}_{Matching} (\matrix{E}),
\end{equation}

\subsection{Multi-Scenario Nearline Retrieval}
In this work, we emphasize the significance of utilizing the ranking results from other scenarios to enhance the matching stage of the current scenario. Specifically, a user may be involved in multiple scenarios $(\vec{s_1}, \vec{s_2}, \cdots, \vec{s_l})$ in a platform. In each scenario $\vec{s}$, the ranking stage will provide a set of scored items based on the ranking model. 
Taking into account that a user may visit a scenario multiple times, the ranking results for a user within scenario $\vec{s}$ can be represented as follows:
\begin{equation}
\begin{split}
\matrix{R}^{t1}_{s} &= ( \vec{e}_1^{t1}, \vec{e}_2^{t1}, \cdots, \vec{e}_n^{t1}), \\
\matrix{R}^{t2}_{s} &= ( \vec{e}_1^{t2}, \vec{e}_2^{t2}, \cdots, \vec{e}_n^{t2}),
\end{split}
\end{equation}
where $t1$ and $t2$ denote different access times of the user, $n$ denotes the number of scoring candidate sets for ranking stage. For each scenario $\vec{s}$, the goal of MNR can be formalized as follow, 
\begin{equation}
\mathcal{E}_{s} = \mathcal{F}_{NMR} (\matrix{R}^{t1}_{s}, \matrix{R}^{t2}_{s}, \cdots, \matrix{R}^{tN}_{s}),
\end{equation}
where $\mathcal{F}_{NMR}$ denotes the function that identifies the top $m$ items critical to the current scenario from various refined ranking candidate sets. $\mathcal{E}_{s} = \{ e_1, e_2, \cdots, e_m \}$ comprises the ultimate selection of items retrieved based on the ranking results within scenario $\vec{s}$.

\paratitle{Scenario-aware Ranking Results Collecting.}
In each scenario, items filtered by the pre-ranking stage are then scored in the ranking stage. Despite obtaining scores for all items during the ranking stage, not every item in the candidate set aligns with user preferences. Consequently, we initially truncate the ranking results obtained from each visit,
\begin{equation}
    \mathcal{R}^{t1}_{s} = Truncate_{s}(\matrix{R}^{t1}_{s}).
\end{equation}
Then it is non-trivial to aggregate several ranking results at different access times (\ie $\mathcal{R}^{t1}_{s}, \mathcal{R}^{t2}_{s}, \cdots, \mathcal{R}^{tl}_{s}$) into a unified set. With online storage being a limiting factor, we utilize a priority queue for each scenario $\vec{s}$ to maintain a record of each user's ranking history,
\begin{equation}
    \mathcal{C}_{s} = Queue(\mathcal{R}^{t1}_{s}, \mathcal{R}^{t2}_{s}, \cdots, \mathcal{R}^{tl}_{s}),
\end{equation}
here we adopt a first-in-first-out (FIFO) strategy to retain the user's latest preferences. In this way, we can retain user interests from other scenarios without incurring extra computational overhead.

\paratitle{Streaming Candidate Scoring.}
Upon compiling the aggregated ranking results $\mathcal{C}{s}$ that reflect user preferences in scenario $\vec{s}$, it's crucial to refine and limit $\mathcal{C}{s}$ to manage the size of the retrieval set. We prioritize two aspects: the initial order of ranking results and the timing of user interaction. The ranking phase typically employs sophisticated models with robust feature interactions, suggesting that items with higher scores are more attuned to user interests. On the other hand, as user tastes can evolve, rankings closer to the present moment often correlate better with current preferences. To reconcile ranking with access timing, we can define the scoring function as follows,
\begin{equation}
    finalScore = (\frac{\alpha}{\alpha + rank\_index}) \times (\frac{\beta}{\beta + time\_index}),
\label{equation:score}
\end{equation}
where $rank\_index$ indicates an item's position in the initial ranking results, arranged in descending order. The term $time\_index$ refers to the sequence of a user's visits, with the index of the most recent visit being $0$. $\alpha$ and $\beta$ act as balancing hyper-parameters, influencing the significance of item ranking and timing of access, respectively. By assigning a higher value to $\alpha$, the scoring formula places greater importance on the timing of the user's visit. Alternatively, elevating $\beta$ steers the focus towards the original ranking score of the items. Based on Equation \ref{equation:score}, the final retrieval results $\mathcal{E}_{s}$ is obtained by by choosing the top-k items from $\mathcal{C}{s}$.

\paratitle{Implementation Details.}
The proposed framework employs the Flink platform for subscribing to rank logs from different scenarios (\ie records of candidate items and their scoring during ranking processes). Quasi-real-time operations such as queue management, aggregation, and top-k calculations are performed. To promote category diversity, we disperse items across categories when generating TopK results, limiting the number of items per category to a predefined threshold. 
% Consequently, our framework relies exclusively on rank logs from various scenarios, eliminating the need for any offline model training or alignment. 
% This approach, in comparison to multi-scenario models, enhances the stability and reproducibility of our framework.

\section{Experiment}
% To comprehensively assess the effectiveness of MNR, we conduct extensive experiments on the Taobao homepage, one of the world's largest traffic recommendation scenarios.

\subsection{Experiment Setting.}
\paratitle{Experiment Scenarios.}
In the work, we evaluate the effectiveness of MNR on \textbf{Taobao Homepage} and MNR utilizes the ranking results from several other scenarios on Taobao. The involved scenarios are as follows:
\begin{itemize}[leftmargin=*]
    \item \textbf{Taobao Homepage}  is the front page of one of the world's largest e-commerce platforms.
    \item \textbf{Main Search} is the largest e-commerce search system in China.
    \item \textbf{Photo Search} is an image search feature integrated within Taobao, enabling users to search for products using images.
    \item \textbf{Post-purchase} activates with recommendations after a user makes a purchase, often spotlighting items that complement the recent buy.
    \item \textbf{In-shop} takes into account the user's interactions within the shop and the shop's inventory to suggest items.
    \item \textbf{Mini-Detail} is a high-traffic scenario on Taobao, presenting users with additional product recommendations alongside the currently viewed item after a click-through.
\end{itemize}

\paratitle{Baselines.}
To further evaluate the effectiveness of our proposed framework, we compare our method with three representative methods in industrial recommendation systems as follows,

\begin{itemize} [leftmargin=*]
    \item \textbf{SWING~\cite{sarwar2001item}} emerges as a state-of-the-art approach in collaborative filtering. It generates recommendations by directly learning a sparse similarity matrix over items and operates by optimizing a regularized objective function.
    \item \textbf{MIND~\cite{li2019multi}} is a deep learning-based algorithm specifically tailored for the recommendation system domain. It uses a dynamic routing mechanism to further enhance its capability to adaptively aggregate these different interests based on the target item.
    \item \textbf{Glare~\cite{amatriain2013system}} operates as a nearline retrieval framework, focusing solely on the model's capabilities within its own scenario, with an implementation that unfolds in a non-real-time manner.
\end{itemize}
% To ensure fair comparison, the same ranking procedure followed the deployment of all methods employed during the matching stage.

\paratitle{Evaluation Metric.}
We adopt two indicators that the industrial recommendation systems pay the most attention to,
\begin{itemize} [leftmargin=*]
    \item \textbf{CTCVR} signifies the likelihood of an item transitioning from exposure to a transaction.
    \item \textbf{Hitrate} indicates the percentage of a user's actual behaviors at all scenarios accurately hit by the items retrieved, which have been normalized based on the number of retrievals.
\end{itemize}
We also present the \textbf{PVR} in the experiment, which represents the percentage of the finally exposed items that come from this particular matching approach. A higher PVR indicates that the following stages have a greater acceptance of the items retrieved.

\subsection{Overall Performance}

\begin{table}[t]
    \centering
    \small
    \caption{The performance of evaluated methods on Taobao Homepage.*}
    \vspace{-0.4cm}
    \begin{tabular}{ccccc}
    \toprule
        ~ & Truncation &  PVR &  Hitrate   & CTCVR \\ \hline
        $Online_{ALL}$ & / & 100\% &  0.62\% & base \\ 
        $SWING$ & / & - &  0.43\% & +21.2\% \\ 
        $MIND$ & / & - &  1.21\% & +7.5\% \\ 
        $Glare$ & / & - &  0.66\% & +0.2\% \\ 
        \hline
        $MNR_{New-Detail}$& 500 & 15.54\%   & 1.94\% & +35.0\% \\ 
        $MNR_{Post-Purchase}$& 500  &  15.29\%   & 1.08\% & +11.3\% \\ 
        $MNR_{MainSearch}$ & 500 & 6.99\% &  2.02\% & +57.0\% \\ 
        $MNR_{In-shop}$ & 20 & 0.71\% & 0.90\% & -33.0\% \\ 
        $MNR_{PhotoSearch}$& 20  & 0.36\%  &  1.66\% & +103.7\% \\ 
        % $MNR_{ALL}$ & / & 0.72\%    &  \% & +93.0\% \\ 
    \bottomrule
    \end{tabular}
    \begin{tablenotes}
        \footnotesize
        \item[a] \textbf{*}Due to data confidentiality requirements, we only display relative values for CTCVR and do not reveal the PVR of SWING, MIND and Glare.
    \end{tablenotes}
    \vspace{-0.2cm}
    \label{table:overall_result}
\end{table}

Table \ref{table:overall_result} illustrates the performance outcomes associated with different methods, where $Online_{ALL}$ represents the overall performance at the matching stage.
Noteworthy in the table is the robust performance uplift observed in $MNR_{MiniDetail}$ and $MNR_{MainSearch}$, marking improvements of 35.0\% and 57.0\% in CTCVR, respectively.  
These metrics underscore the efficacy of MNR's strategic utilization of cross-scenario insights, elevating its precision in predicting and matching user preference.
In terms of Hitrate, $MNR_{MiniDetail}$ and $MNR_{MainSearch}$ report substantial improvements, with Hitrate of 1.94\% and 2.02\%, outperforming the foundational 0.62\% of $Online_{ALL}$.
Besides, the results from Table \ref{table:overall_result} further indicate that MNR also surpasses other competing methods such as SWING, MIND, and Glare. It can be observed that MNR represents significant improvements over Swing's 0.43\%, Mind's 1.21\%, and Glare's 0.66\%, respectively, translating to more accurately targeted recommendations.
Moreover, scenarios like $MNR_{In-Shop}$ and $MNR_{PhotoSearch}$ are challenged by a lack of ample finely ranked scores, leading to the implementation of a lower truncation threshold. Consequently, this curtails the quantity of items funneled into the matching pool, resulting in a diminished PVR.
The underwhelming performance of $MNR_{In-Shop}$ can be linked to the nature of in-shop scenarios, where users are limited to the items of a specific shop once they enter it, typically ranging from a few dozen to about a hundred items. The majority of products are likely to be exposed, and so those captured by the MNR framework for recall are predominantly tail-end items.

\subsection{Ablation Study}

\begin{table}[t]
    \centering
    \small
    \caption{The performance of evaluated methods.*}
    \vspace{-0.4cm}
    \begin{tabular}{cccc}
    \toprule
        ~& Aggregation & PVR    & CTCVR \\ \hline
        $MNR_{Main Search}$ & Online Streaming & 6.99\% & base \\ 
        $MNR_{Main Search}$ & Offline Batch & 3.02\%  & -36.0\% \\ 
    \bottomrule
    \end{tabular}
    \begin{tablenotes}
        \footnotesize
        \item[a] \textbf{*} We conduct our ablation study using $MNR_{MainSearch}$, as MainSearch represents the scenario with the highest traffic on Taobao.
    \end{tablenotes}
    \vspace{-0.2cm}
    \label{table:online}
\end{table}

\paratitle{Impact of Online Stream Computing.}
To evaluate the impact of online stream computing, we compare the MNR with online~(\ie real-time stream computing) and offline~(\ie daily updated) modes. From Table \ref{table:online},  MNR with  offline mode shows a significant reduction in performance with a 36.0\% decline, registering a PVR of just 3.02\%. This stark contrast highlights the critical advantage of real-time data processing offered by the online mode, cementing the importance of stream computing in enhancing the accuracy and efficiency of recommendation models.

\begin{table}[t]
    \centering
    \small
    \caption{The performance of NRM compares with FIFO strategy and random strategy.}
    \vspace{-0.4cm}
    \begin{tabular}{cccc}
    \toprule
        ~ & Strategy & PVR   & CTCVR \\ \hline
        $MNR_{MainSearch}$ & Streaming Candidate Scoring & 6.99\% & base \\ 
        $MNR_{MainSearch}$ & Only FIFO & 1.93\%  & -8.9\% \\ 
        $MNR_{MainSearch}$ & Random & 0.35\%  & -23.6\% \\ 
    \bottomrule
    \end{tabular}
    \vspace{-0.4cm}
    \label{table:fifo}
\end{table}

\paratitle{Impact of Streaming Candidate Scoring.}
MNR utilizes Equation \ref{equation:score} to score the ranking results collected in the queue from each scenario. To validate the effectiveness of our proposed scoring function, we evaluate against both a First-In-First-Out (FIFO) strategy as well as a random strategy, to evaluate their impacts.
From Table \ref{table:fifo}, we can observe that MNR shows a significant improvement in effectiveness when compared to alternative strategies. The base PVR of MNR stands at 6.99\%. In contrast, the MNR with FIFO strategy, focusing solely on the temporal aspect and ignoring the ranking scores, shows a substantial decrease in performance with only a 1.93\% effectiveness rate and an 8.9\% reduction compared to the base. Similarly, the MNR with random strategy, which employs a completely random selection method, results in an even more pronounced drop to 0.35\% effectiveness, with a 23.6\% reduction. This comparison highlights the superior efficacy of the MNR's targeted scoring approach in optimizing recommendation outcomes over simpler FIFO or random strategies.

\begin{table}[t]
    \centering
    \small
    \caption{The impact of hyper-parameter $\alpha$.}
    \vspace{-0.4cm}
    \begin{tabular}{cccc}
    \toprule
        & $\alpha$ & PVR   & CTCVR \\ \hline
        $MNR_{MainSearch}$ & 10 & 3.71\% & -2.0\% \\ 
        $MNR_{MainSearch}$ & 20 & 3.65\%  & -1.7\% \\ 
        $MNR_{MainSearch}$ & 50 & 3.71\%  & base \\ 
        $MNR_{MainSearch}$ & 200 & 3.63\%  & -12.0\% \\ 
        $MNR_{MainSearch}$ & 500 & 3.57\%  & -15.0\% \\ 
    \bottomrule
    \end{tabular}
    \vspace{-0.5cm}
    \label{table:alpha}
\end{table}

\paratitle{Impact of hyper-parameter $\alpha$.}
$\alpha$ plays a crucial role as a trade-off hyper-parameter, mediating the balance between ranking influence and access time. 
The results of the impact of hyper-parameter $\alpha$ are shown in Table \ref{table:alpha}. Notably, as $\alpha$ increases, we observe a trend where PVR remains relatively stable but CTCVR experiences a significant decline. At $\alpha$ values of 200 and 500, the CTCVR drops by -12.0\% and -15.0\%, respectively, indicating that setting $\alpha$ too high might lead to an overemphasis on visit time at the expense of relevant ranking factors. This trend highlights the importance of carefully calibrating $\alpha$ to strike an optimal balance between recency and relevance in our MNR framework.

\subsection{Online performance.}
We conducted online experiments by deploying MNR with multiple selected scenarios data to handle the real traffic of Taobao's homepage for a month. 
The results from real traffic on the Taobao homepage demonstrate that our proposed method, compared to the baseline, achieves a consistent and stable increase of 5\% in transaction volume. This highlights the commercial value and practicality of our approach. Moreover, the proposed algorithm exhibits efficient computational performance and low resource consumption, rendering it a promising solution for real-time personalized recommendation scenarios in recommendation systems.

\section{conclusion}

In conclusion, the proposed multi-scenario nearline retrieval framework enhances the matching stage of recommendation systems, addressing the constraints of scale and latency by leveraging ranking logs from various scenarios in near real-time. The integration of this framework within the Taobao APP has shown a commendable increase in product transactions by 5\%, indicating its effectiveness and potential for broader application. This advancement underscores the benefits of cross-scenario information utilization and in future work, we intend to optimize our pre-ranking model by integrating recalls from multiple scenarios.

%%
%% The next two lines define the bibliography style to be used, and
%% the bibliography file.
\bibliographystyle{ACM-Reference-Format}
\bibliography{sample-base}

\end{document}